\documentclass{report}
 \usepackage[ansinew]{inputenc}
\usepackage[]{epsfig}

\begin{document}

\setlength{\textheight}{20cm}
\setlength{\textwidth}{13.5cm}
\renewcommand{\abstract}[1]{{ \footnotesize \noindent {\bf Abstract:} #1 \\}}
\renewcommand{\author}[1]{\subsubsection*{\it#1}}
\newcommand{\address}[1]{\subsubsection*{\it#1}}
\newcommand{\ltsimaZaroubi}{$\; \buildrel < \over \sim \;$}
\newcommand{\lsimZaroubi}{\lower.5ex\hbox{\ltsimaZaroubi}}
\newcommand{\gtsimaZaroubi}{$\; \buildrel > \over \sim \;$}
\newcommand{\gsimZaroubi}{\lower.5ex\hbox{\gtsimaZaroubi}}

\chapter*{Cosmic Flows: Review
of Recent Developments}

\author{Saleem Zaroubi}
\address{Max Planck Institute for Astrophysics\\
Karl-Schwarzschild-Str. 1\\
 D-85741 Garching, Germany
}

\abstract{I review the recent developments in the analysis of cosmic
flow data, in particular, latest results of bulk flow measurements,
comparison between redshift and peculiar velocity catalogs with
emphasis on the measured value of the $\beta\, (=\Omega_m^{0.6}/b)$
parameter, and matter power spectrum estimates from galaxy peculiar
velocity catalogs. Based on these developments, one can argue that
most of the previous discrepancies in the interpretation of cosmic
flow data, {\it maybe} with the exception of bulk flow measurements on
scales $\gsimZaroubi 100 h{^{-1}}{\rm Mpc}$, have either been resolved or
fairly understood.  }
 
\section{Introduction}
\label{sec:zaroubi_introduction}
Within the gravitational instability (GI) framework for the growth of
cosmic structures, the peculiar velocity field of galaxies and
clusters provides a direct and reliable probe of the matter
distribution, under the natural assumption that these objects are
unbiased tracers of the large-scale, gravitationally induced, velocity
field. The GI paradigm requires that the linear peculiar velocity,
${\bf v}({\bf r})$ -- defined as the deviation from Hubble expansion --
and linear mass density contrast, $\delta_m({\bf r})$, be related to
one another according to the local (differential) relation,
\begin{equation} 
\nabla \cdot {\bf v} = -\Omega_m^{0.6} \delta_m,
\label{eq:zaroubi_divv}
\end{equation}
or its global (integral) counterpart,
\begin{equation} 
{\bf v} = {\Omega_m^{0.6} \over 4\pi} \int d^3{\bf r'} { \delta_m({\bf
r'}) ({\bf r'}-{\bf r}) \over \vert {\bf r'}-{\bf r}\vert^3},
\label{eq:zaroubi_intd}
\end{equation}
where $\Omega_m$ is the matter overdensity parameter.
Note that the peculiar velocity field is determined by the
distribution of the matter with all its components especially the
dominant dark matter component.

In order to measure peculiar velocities of galaxies and clusters,
observers use a variety of distance indicators. Generally, these
indicators relate two quantities, one among which is distance
dependent, {\it e.g.,} galaxy luminosity, and the other is distance
independent, {\it e.g.,} galaxy rotational velocity. The best known
examples of such indicators are the Tully-Fisher~\cite{tully77} and
Faber-Jackson~\cite{faber76} relations; over the last decade or so
these and many other types of distance indicators have been used to
measure cosmological distances. The availability of a large number of
galaxy peculiar velocity catalogs, some of them with few thousands
objects, have turned cosmic flows to one of the main probes used to
study the large scale structure in the nearby universe. Here I'll
concentrate on the following three statistical measures of the
velocity field:

\begin{enumerate}
\item {\it The Bulk Flow:} This measure, defined as the average
streaming motion within certain volume, is probably the easiest
statistic to estimate from the observed radial component of peculiar
velocities. At the Cosmic Microwave Background radiation (CMB)
restframe, the bulk motions are expected to converge to zero with
increasing volume. The rate of convergence depends on the fluctuations
in the matter distribution on various scales, {\it i.e.,} the large
scale matter fluctuations power spectrum. This dependence on
cosmological models has motivated several attempts to measure the
dipole component of the local peculiar velocity field and to determine
the volume within which the streaming motion vanishes. As of yet,
bulk flow measurements have produced conclusive and consistent results
only on scales $\lsimZaroubi 60h{^{-1}}{\rm Mpc}$, but failed to do so on scales
$\gsimZaroubi 100 h{^{-1}}{\rm Mpc}$ (for recent works see \cite{colless01,dacosta00,
courteau00, dale99, giovanelli97, hudson99, willick99}).

\item
{\it The mass power spectrum:} Equation~\ref{eq:zaroubi_intd} suggests that
one can estimate the bias free, $\Omega_m^{1.2}$ weighted, matter
power spectrum directly from the measured peculiar velocities. To date,
likelihood analysis based estimations of the matter power spectrum
exist for the Mark~III~\cite{zaroubi97}, SFI~\cite{freudling99} and
ENEAR~\cite{zaroubi01} catalogs. All of these
measurements has consistently produced power spectra with amplitudes
larger than those measured by other data sets, {\it e.g.,} galaxy redshift
surveys.

\item
{\it The $\beta$ parameter:} Peculiar velocities enable a
reconstruction of the large scale matter distribution independent of
redshift surveys. Therefore, one can use Eqs.~\ref{eq:zaroubi_divv}
and~\ref{eq:zaroubi_intd} to compare the matter and 3D
velocity distributions deduced from the measured radial peculiar
velocities to those obtained from redshift surveys. This comparison
requires biasing model which specifies how galaxies follow the
underlying {\it total} matter distribution. On the scales of interest,
it is usually assumed that 
\begin{equation}
\delta_g = b \delta_m ,
\label{eq:zaroubi_bias}
\end{equation}
where $\delta_g$
is the galaxy observed density contrast and $b$ is the linear bias
parameter.  The comparison is used to: 1) test the validity of
Eqs.~\ref{eq:zaroubi_divv} and~\ref{eq:zaroubi_intd}, namely, the
basic GI paradigm; 2) test the linear biasing model; 3) directly
measure the value of $\Omega_m$ (\cite{branchini01, dacosta98,
davis96, sigad98, willick98, willick97b} and \cite{zaroubi02b}). Until
recently, comparisons using Eq.~\ref{eq:zaroubi_divv} have
systematically yielded $\beta$ values larger than those obtained from
using Eq.~\ref{eq:zaroubi_intd}.
\end{enumerate}

As mentioned above, many have attempted to estimate these statistical
measures of peculiar velocities over the years. Until recently, the
results they obtained have often been inconsistent with each other and
with estimations from other data, depending on the specific peculiar
velocity catalog at hand and the analysis methods.  It is generally
accepted that the main reason for the inconsistencies lies in the
problematic nature of the distance indicators used to determine the
peculiar velocities. First, the distance measurements carry large
random errors, including intrinsic scatter in the distance indicator
and measurement errors, which grow in proportion to the distance from
the observer and thus become severe at large distances. Further
nontrivial errors are introduced by the nonuniform sampling of the
galaxies that serve as velocity tracers. In particular, the Galactic
disk obscures an appreciable fraction of the sky, creating a
significant``zone of avoidance'' of at least $40\%$ of the sky. When
translated to an underlying smoothed field, these errors give rise to
severe systematic biases. In light of these difficulties, the
inconclusive results have led many to question the reliability of the
peculiar velocity datasets as cosmological probes.

\begin{figure}
    \vskip -1.5 truecm
  \centering \begin{tabular}{c} \hskip -1 truecm
    \mbox{\epsfig{file=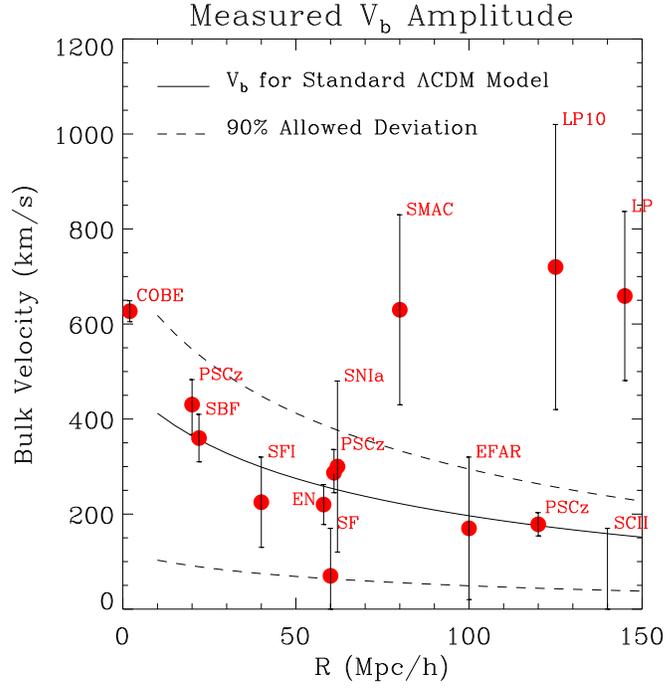,height=9cm}} \\
    \mbox{\epsfig{file=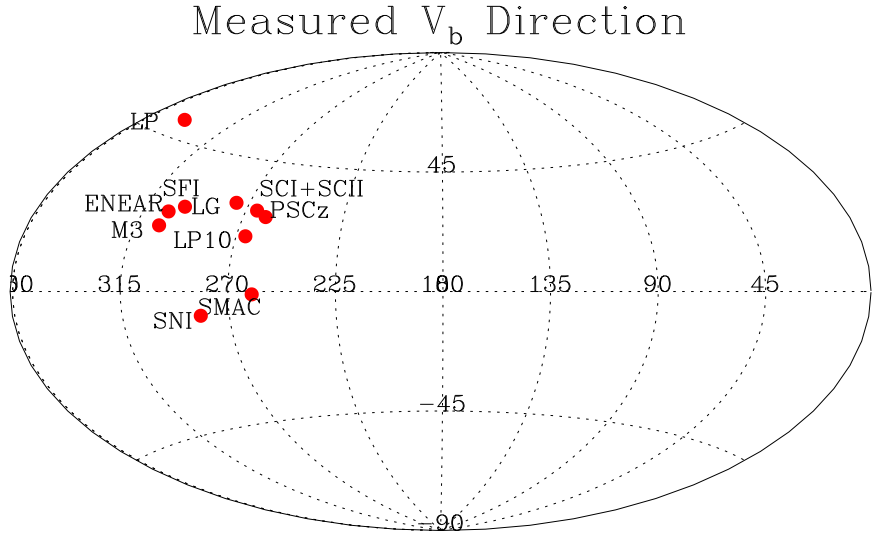,height=7cm}}
    \end{tabular} 
    \vskip 1 truecm
    \caption{{\small Bulk Flow measurements. Upper panel:
    the symbols show the amplitude of the measured bulk flow (with its
    error) from the following surveys: Surface Brightness Fluctuations
    (SBF), SFI , ENEAR (EN), Shellflow (SF), Supernovae type Ia
    (SNIa), SMAC, EFAR, LP10, SCII and LP (see table for explanation)
    as a function of radius. The CMB dipole COBE measurement and bulk
    flow from the PSC$z$ redshift catalog are also shown. The solid
    line shows the expected rms bulk velocity of a sphere of radius
    $R$ for standard $\Lambda$CDM model; the dashed lines represent
    1-$\sigma$ cosmic scatter about the rms.  Lower panel: the symbols
    show the direction of some of the measured bulk flow vectors,
    note that the catalogs that correspond to $R\sim 60 h{^{-1}}{\rm
    Mpc}$ have consistent directions while measurements that
    correspond to large distances do not. }} \label{fig:zaroubi_fig1}
\end{figure}
In this article, I review the recent developments in these three areas
and show that significant improvements have occurred in the field in
the past few years. In addition, an argument is put forward that those
developements lead to alleviating most of the inconsistencies and
to understanding, at least qualitatively, the cause of the remaining
outstanding ones.

\section{Bulk Flow}
\label{sec:zaroubi_bulk}
The bulk flow motion direction and amplitude on different scales are
the simplest quantities to measure from peculiar velocity data. They
provide constraints on the power-spectrum of mass
fluctuations. Theoretically, the mean square bulk velocity within a
sphere of radius $R$, is given by,
\begin{equation}
\langle v^2(R)\rangle = {\Omega_m^{1.2} \over 2 \pi^2} \int_0^\infty
P(k) \tilde W^2(kR) dk,
\label{eq:zaroubi_bulk}
\end{equation} 
where $P(k)$ is the mass fluctuation power spectrum and $\tilde W(kR)$
is the Fourier transform of a top-hat window of radius R. In the upper
panel of Figure~\ref{fig:zaroubi_fig1} the rms expected bulk velocity,
${V_b}^{rms}(R) = \sqrt{\langle v^2(R)\rangle},$ is plotted against R
for a standard $\Lambda$CDM power spectrum with the dashed lines
representing the $1\sigma$ cosmic scatter deviations expected within the
given volume. Note that the large cosmic scatter weakens the bulk flow
statistic as a probe of cosmological models.

On the observational side one can divide the currently available
measurements into two main domains. The first is measurements of ${\bf
V}_b$ within radius $\lsimZaroubi 60h ^{-1} Mpc$. Bulk flow values
within this radius as measured from the SFI~\cite{dacosta96},
Mark~III~\cite{dekel99} and the recently completed
Shellflow~\cite{courteau00} and ENEAR~\cite{dacosta00} catalogs, lead
to a roughly consistent picture even though some discrepancies still
remain. For instance, the Mark~III catalog yields a systematically
larger amplitude of the bulk motion $\sim370$~${\rm km\,s{^{-1}}}$\ on
scales $\sim 60 h{^{-1}}{\rm Mpc}$ as compared to values of
$\lsimZaroubi 220$~${\rm km\,s{^{-1}}}$ obtained from the other
catalogs; this specific discrepancy is probably caused by the
calibration problem the Mark~III catalog is known to suffer
from~\cite{courteau00, davis96}.

Recently, results from the long awaited Surface Brightness
Fluctuations (SBF) method have been first
published~\cite{tonry00}. They give a bulk flow value of $350 {\rm
km\,s{^{-1}}}$ at $30h{^{-1}}{\rm Mpc}$ distance.  The bulk flow
measurement within this volume, both in terms of amplitude and
direction, is consistent with what is known about the structures just
outside the sampled volume, {\it e.g.,} the Great-Attractor the
Perseus-Pisces superclusters.

The second domain is measurements within $R \gsimZaroubi 100
h{^{-1}}{\rm Mpc}$.  The results here are far from being conclusive
and various data sets lead to different bulk flow amplitudes and
directions. While there is supporting evidence pointing towards
convergence of the bulk flow on these large scales~\cite{dale99,
riess97, colless01}, other works~\cite{lauer94, willick99, hudson99}
argue for the existence of large amplitude ($\gsimZaroubi 600$~${\rm
km\,s{^{-1}}}$) streaming motions out to a depth as large as $150
h{^{-1}}{\rm Mpc}$, ruling out that the Hubble flow has converged to
the CMB frame at smaller distances.  Given the far reaching
implications that these large-scale motions would have on currently
popular cosmological models it is clear that this issue is of great
interest. It is important to point out, however, that the direction
and the amplitude of the bulk motion detected on large scales by
different authors do not agree with each other. Additional independent
bulk flow estimates, {\it e.g.,} from the PSC$z$ redshift catalog and
more recently from the dipole of the NVSS radio
sources~\cite{blake02}, are consistent with bulk flow convergence in
the CMB restframe (see upper panel of Fig.~\ref{fig:zaroubi_fig1}).

A summary of the bulk flow measurements is given in
Fig.~\ref{fig:zaroubi_fig1} and Table~\ref{table:zaroubi_table1}.

\begin{table}
\begin{center}
\caption{Recent Bulk Flow Measurements \label{table:zaroubi_table1}}
\begin{tabular}{lccc}
\hline
\hline
Survey    & $V_b$ & R & Comments \\
 &  (${\rm km\,s{^{-1}}}$ ) & ($h^{-1}{\rm Mpc}$ ) & \\
\hline

Tonry {\it et al.}  \cite{tonry00}  & $350$  & $30$ & SBF$^\ast$ \\
\hline
Dekel {\it et al.}  \cite{dekel99}  & $370$  & $60$ & M3$^\star$ (TF$^\dagger$ + {$D_n-\sigma$}) \\
Giovanelli {\it et al.}   \cite{giovanelli97} &   $200$ & $60$ & SFI (TF) \\
Courteau {\it et al.}  \cite{courteau00} &  $70$ & $60$ & Shellflow (TF) \\
da Costa {\it et al.}  \cite{dacosta00}&    $220$ & $60$ & ENEAR ({$D_n-\sigma$}) \\   
\hline
Riess {\it et al.}  \cite{riess97} & $\approx 0$ & $100$ & SNIa$^\bullet$ \\ 
Colless {\it et al.}  \cite{colless01} & $\approx 0$ & $\sim 100 $ & EFAR (FP$^\diamond$) \\
Hudson {\it et al.}  \cite{hudson99} & $600$ & $140$ & SMAC (FP) \\
Lauer \& Postman \cite{lauer94}   &  $700$ & $150$ & LP (BCG$^\ddagger$) \\
Willick \cite{willick99}  & $700$  & $150$ & LP10K (TF)  \\
Dale {\it et al.}   \cite{dale99} & $\approx 0$ & $\sim 150$ & SCI/SCII (TF) \\
\hline
\end{tabular}
\end{center}
{\footnotesize  \noindent $^\ast$~Surface Brightness Fluctuations method.  
                  	  $^\star$~Mark~III dataset.
                  	  $^\dagger$~Tully-Fisher Measurement.
                  	  $^\bullet$~Supernovae type Ia.
                  	  $^\diamond$~Fundamental Plane measurement.
                  	  $^\ddagger$~Brightest Cluster Galaxy method. }
\end{table}

\section{Power Spectrum Analysis}
\label{sec:zaroubi_ps}

\begin{figure}
  \centering \epsfig{file=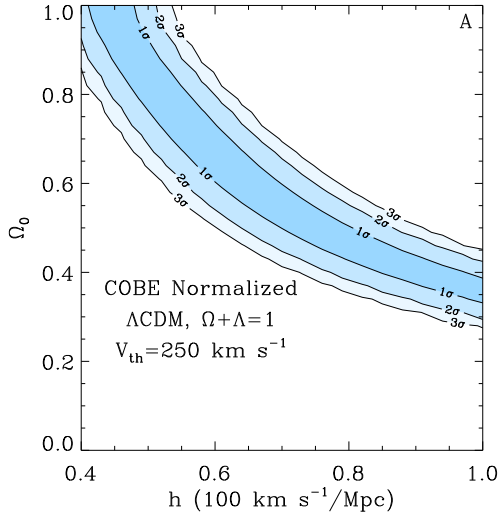,height=7.5cm}
  \caption{Contour map of ln-likelihood in the $h -\Omega$ plane for
  $\Lambda$CDM models with $250$ ${\rm km\,s{^{-1}}}$ thermal error component (added
  to account for the nonlinear environment in which early type
  galaxies tend to reside). The contours denote the most likely values
  within $1,2$ and $3\sigma$ confidence levels.  }
  \label{fig:zaroubi_fig2}
\end{figure}

The matter power spectrum has been measured from the Mark~III
\cite{kolatt97, zaroubi97}, SFI \cite{freudling99} and ENEAR
\cite{zaroubi01} catalogs.  With the exception of one
study~\cite{kolatt97}, all the power spectrum estimations apply
likelihood analysis which assumes that both the underlying velocity
field and the errors are drawn from independent random Gaussian
fields. The observed peculiar velocities constitute a multi-variant
Gaussian data set, albeit the sparse and inhomogeneous sampling; this
probability distribution function is reinterpreted as a likelihood
function of the measured radial velocities given a model power
spectrum. Maximizing the likelihood with respect to the model free
parameters yields a best fit power spectrum.


Together with its advantages (no explicit window function, weighting
or smoothing the data and automatic underweighting of noisy,
unreliable data) the likelihood method requires satisfying
the following conditions: 1) explicit knowledge of the distribution
function; 2) peculiar velocities are related to the over-densities
through linear theory; 3) errors in the inferred distances constitute
a Gaussian random field with scatter that scales linearly with
distance. In addition, the likelihood analysis requires assuming a
parametric functional form for the power spectrum.

A typical result of the power spectrum parameters as estimated from,
for example, the ENEAR data is presented in
Fig.~\ref{fig:zaroubi_fig2}.  This figure shows the likelihood contour
map in the $\Omega_m-h$ plane, for the $\Lambda$CDM family of models
with Harrison-Zeldovich spectrum ($n=1$). In addition to the normal,
distance dependent, uncertainties the error matrix is assumed here to
have a diagonal random contribution of $250$ ${\rm km\,s{^{-1}}}$ that
accounts for the nonlinear environment within which the ENEAR early-type
galaxies prefertionaly reside.  The most probable parameters in this
case (in the range $\Omega_m\leq 1$) are $\Omega_m=1$ and $h=0.5$. The
elongated contours clearly indicate that neither $\Omega_m$ nor $h$
are independently well constrained.  It is rather a degenerate
combination of the two parameters, approximately $\Omega_m h^x$ with
$x\sim 1$ that is being determined tightly by the elongated ridge of
high likelihood.

As pointed out earlier, the power spectrum amplitude deduced from
galaxy peculiar velocity data is considerably higher than those
obtained from other types of data sets, usually favor the standard
$\Lambda$CDM model ($\Omega_m=0.3$, $\Lambda=0.7$ and $h=0.65$). This
naturally raises the question of whether is the discrepancy driven by
a yet undetected systematic effect in the data, or by something
deeper?

It is important to point out that the power spectrum estimation method
is very sensitive to the assumed small scale power model (induced by
noise and/or nonlinear evolution) which can add or suppress
power. Recently, Hoffman and Zaroubi~\cite{hoffman00} have carried out
a detailed inspection of the goodness-of-fit of the Mark~III, SFI, and
ENEAR best-fit power spectra by using Principal Component Analysis
(PCA) approach and found that in neither of the three cases do the
assumed theoretical power spectrum and/or error model give an acceptable
fit. In a subsequent study, Silberman {\it et al.}~\cite{silberman01}
have shown that this misfit is probably caused by ignoring small scale
power in the analysis; thereby casting a large shadow of doubt about
the applicability of the error models or the strict validity of linear
dynamic, assumed so far.

\section{The Value of $\beta$}
\label{sec:zaroubi_beta}

\begin{figure}
  \centering \begin{tabular}{l l} \hskip -1.5 truecm
   \mbox{\epsfig{file= 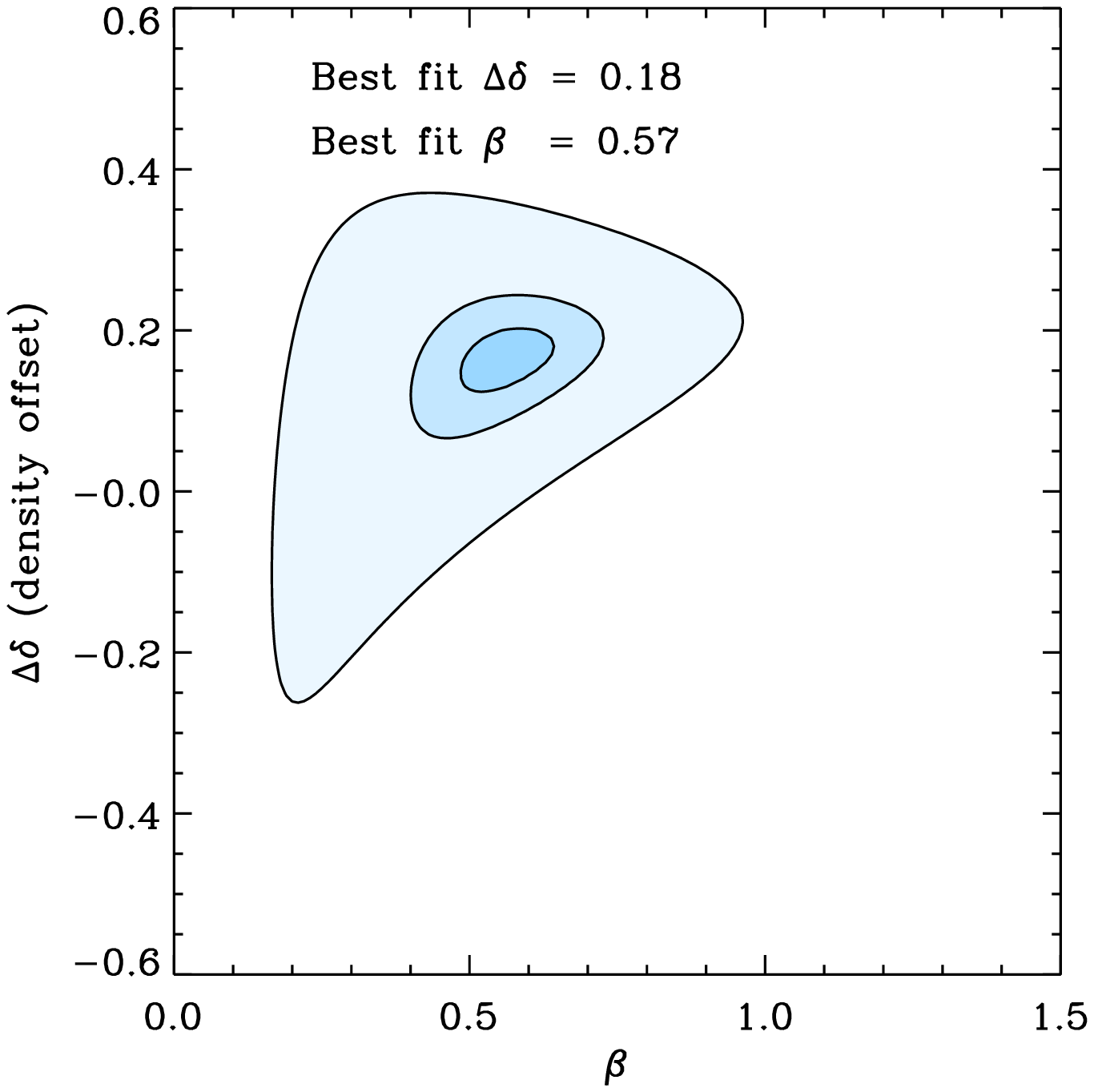,height=7cm}} &
   \mbox{\epsfig{file= 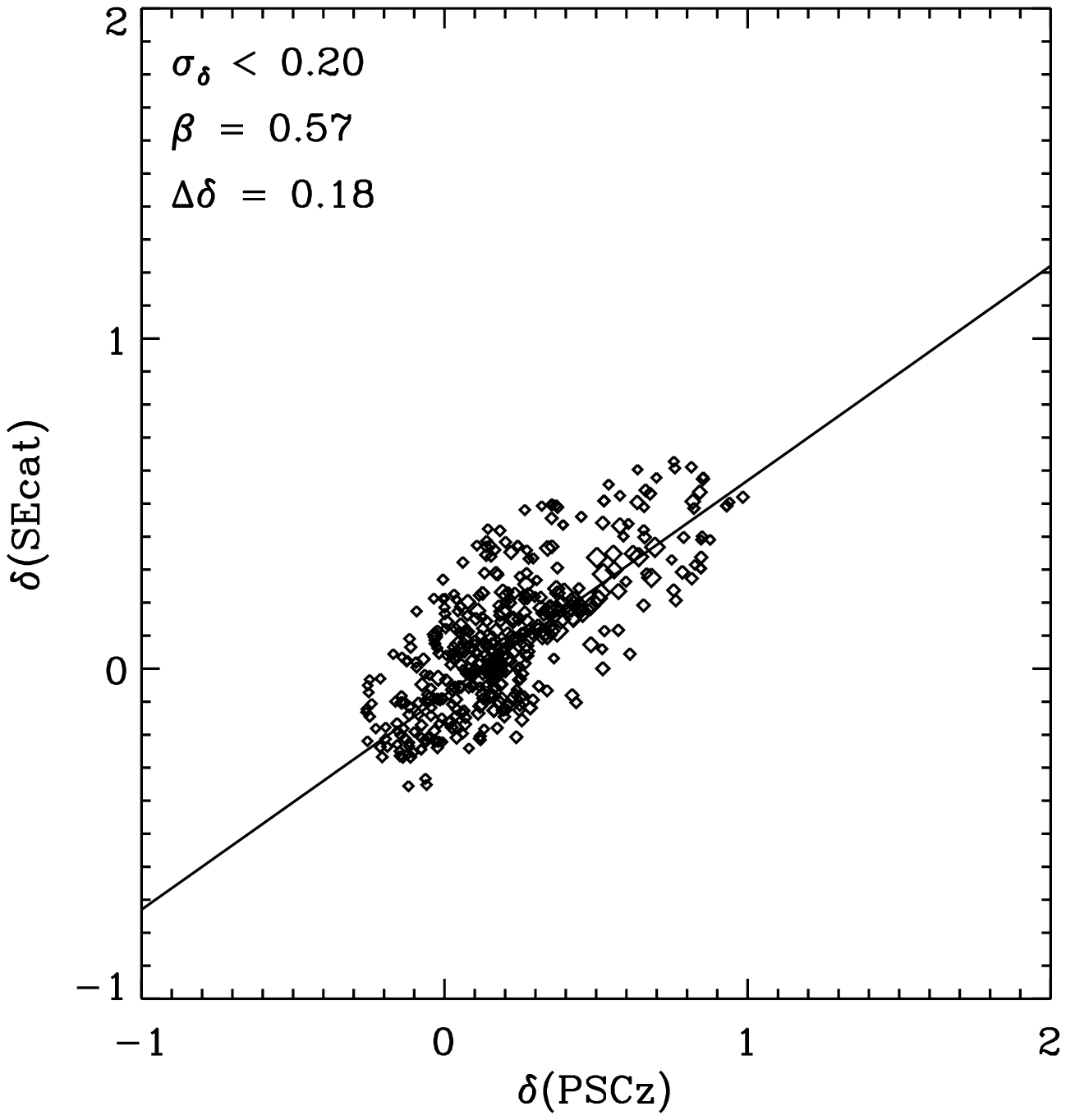,height=7cm}} \end{tabular}
   \caption{\small{$\beta$ from density-density comparison. Left
   Panel: The 1, 2 and 3 $\sigma$ likelihood contours in the $\beta -
   \Delta\delta$ plane. Right Panel: A scatter plot showing the SEcat
   UMV reconstructed G12 density {\it vs.} the PSC$z$ G12 density. In
   this comparison, one out of ten from all the data points that have
   reconstruction error $< 0.2$ was randomly picked. The size of the
   symbols is inversely proportional to their errors. } }
   \label{fig:zaroubi_fig3}
\end{figure}

Galaxy peculiar velocities and their redshift space positions have
been used to estimate the value of $\beta=\Omega_m^{0.6}/b$, under the
hypotheses of linear theory and linear biasing. These
analyses have been typically carried out using two alternative
strategies. In the so-called density-density comparisons a 3-D
velocity field and a self-consistent mass density field are derived
from the observed radial velocities and compared to the galaxy density
field measured from large redshift surveys, under the assumption of
linear bias (a combination of Eqs.~\ref{eq:zaroubi_divv}
\&~\ref{eq:zaroubi_bias}). The typical example here is the comparison
of the density field reconstructed applying the POTENT method
\cite{bertschinger89, dekel90} to the MARK III catalog of galaxy
peculiar velocities \cite{willick97a} with the IRAS 1.2 Jy redshift
catalog density field \cite{sigad98}. The various applications of
density-density comparisons to a number of datasets have persistently
led to large estimates of $\beta$, consistent with unity
\cite{sigad98}. The alternative approach is that of the
velocity-velocity analysis. In this second case the observed galaxy
distribution is used to infer a mass density field from which peculiar
velocities are obtained and compared to the observed ones (a
combination of Eqs.~\ref{eq:zaroubi_intd}
\&~\ref{eq:zaroubi_bias}). The velocity-velocity methods have been
applied to most catalogs presently available yielding systematically
lower values of $\beta$, in the range of $0.4 - 0.6$.

Both density-density and velocity-velocity methods have been carefully
tested using mock catalogs extracted from N-body simulations. They
were shown to provide an unbiased estimate of the $\beta$
parameter. Yet, when applied to the same datasets the discrepancy in
the $\beta$ estimates turned out to be significantly larger than the
expected errors. Velocity-velocity comparisons are generally regarded
as more robust as they require manipulation of redshift catalog data
as opposed to the density-density comparison which manipulates the
noisier and sparser peculiar velocity data. In any case both
approaches are quite complicated and it is hard to understand how
systematic errors can arise and propagate through the analysis. It is,
therefore, likely that these systematic effects do influence the
$\beta$ parameter estimation.
\begin{figure}
  \centering \begin{tabular}{l l} \hskip -1.5 truecm
   \mbox{\epsfig{file= 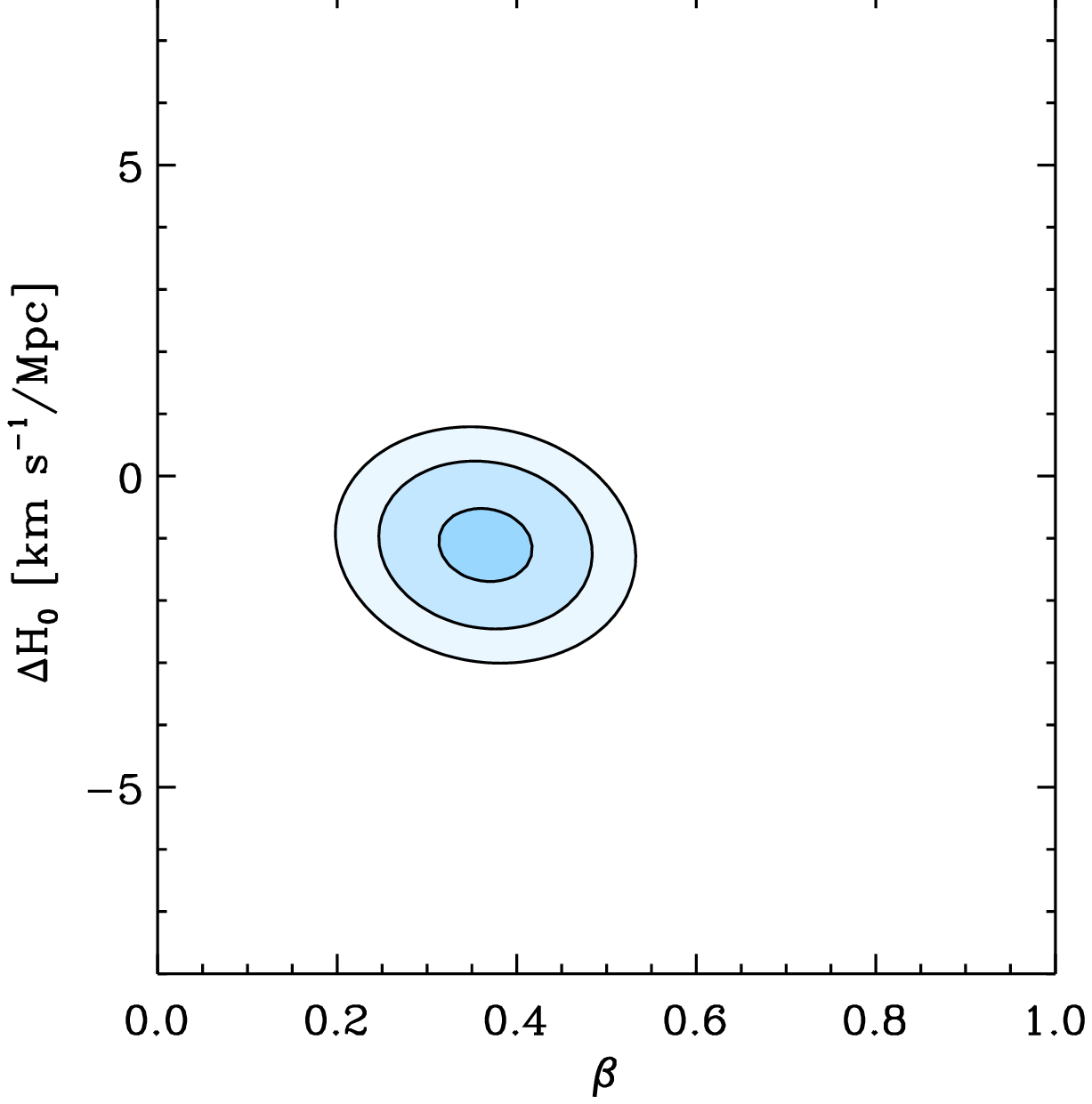,height=7cm}} &
   \mbox{\epsfig{file= 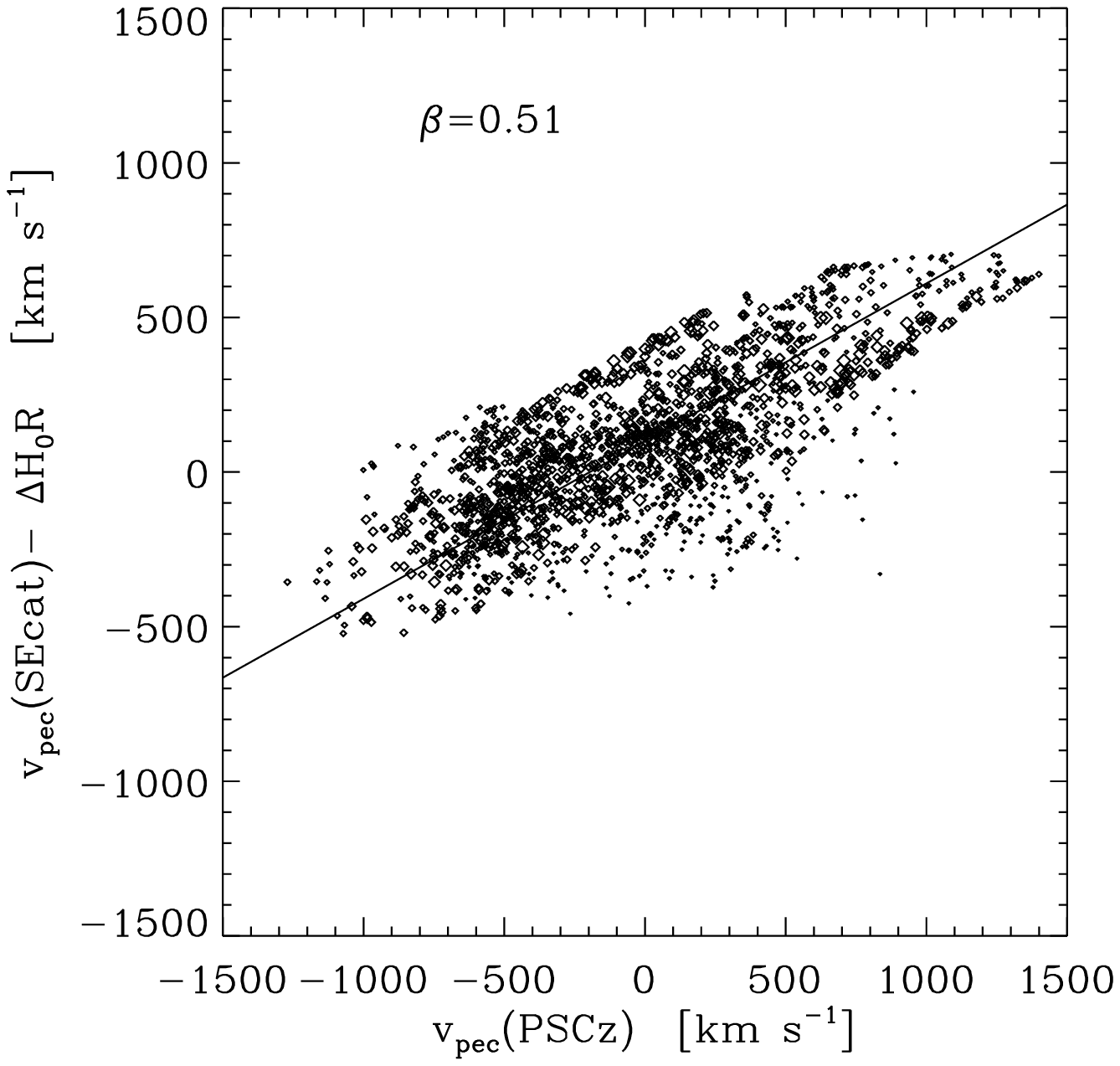,height=7cm}} \end{tabular}
   \caption{{\small $\beta$ from velocity-velocity comparison. Left
   Panel: The 1,2, and 3 $\sigma$ likelihood contours in the
   $\beta-\Delta H_{\circ}$ plane from a comparison of the SEcat
   G12-smoothed radial velocities at the locations of the SEcat data
   points {\it vs.} the PSC$z$ G12-smoothed radial velocities at the
   same locations comparison. Right Panel: A scatter plot of the
   G12-smoothed radial velocities, in this comparison all data points
   have been included. The size of the symbols is inversely
   proportional to the errors }} \label{fig:zaroubi_fig4}
\end{figure}

Since all the previous results have been obtained with methods
designed to conclusively carry out either velocity-velocity comparison
(ITF \cite{davis96} \& VELMOD \cite{willick98, willick97b}) or
density-density comparison (POTENT~\cite{sigad98}), it is very
difficult to know whether the source of the discrepancy is data or
methodology driven. Recently, a new linear approach, called unbiased
minimal variance (UMV) method has been proposed~\cite{zaroubi02a}.
With the UMV method, both velocity-velocity and density-density
comparisons can be carried out within the same methodological
framework.

Similar to the Wiener Filter~\cite{wiener49, zaroubi95}, the UMV
estimator is derived by requiring the linear minimal variance solution
given the data and an assumed {\it prior} model specifying the
underlying field covariance matrix. However, unlike the Wiener filter,
the minimization is carried out with the added constraint of an
unbiased reconstructed mean field.  In the context of reconstruction
from peculiar velocity data, the UMV algorithm could be regarded as a
compromise between the POTENT algorithm, which assumes no
regularization but might be unstable to the inversion problem of
deconvolving highly noisy data, and the Wiener filter algorithm, which
takes into account the correlation between the data points and
therefore stabilizes the inversion, but constitutes a biased estimator
of the underlying field (for detailed discussion see \cite{zaroubi02a,
zaroubi02b}).

The UMV approach has been applied to the SFI, ENEAR, Mark~III and to
the newly compiled SEcat catalog (which is a combination of the SFI
and ENEAR catalogs)~\cite{zaroubi02b}.  The reconstructed fields are
compared with those predicted from the IRAS PSC$z$ galaxy redshift
survey to constrain the value of $\beta$.  For example, the analysis
of the SEcat catalog for the first time leads to consistent $\beta$
values from the comparison of the density and the velocity fields
yielding $\beta=0.57_{-0.13}^{+0.11}$ and $\beta=0.51 \pm{0.06}$,
respectively. The $\beta$ values obtained from comparing the PSC$z$
data to each of the four galaxy peculiar velocity catalogs are consistent.

\begin{table}
\begin{center}
\caption{ $\beta$ Measurements \label{table:zaroubi_table2}}
\begin{tabular}{lcc}
\hline
\hline
Method & Compared data & $\beta$ \\
\hline
\hline
    & $\delta -\delta$ Comparison & \\
\hline
\hline
POTENT \cite{sigad98} & Mark III {\it vs.} IRAS 1.2Jy & $ 0.89 \pm 0.12 $\\
UMV\cite{zaroubi02b} & SEcat {\it vs.} PSCz & $ 0.57^{+0.11}_{-0.13}$ \\

\hline
\hline
 & $v - v$ Comparison & \\
\hline
\hline
VELMOD \cite{willick97b, willick98}  & Mark III {\it vs.} IRAS 1.2Jy & $
           0.50 \pm 007$ \\
VELMOD \cite{branchini01}  & SFI  {\it vs.} PSCz & $ 0.42 \pm 0.07$ \\ 
 ITF   \cite{davis96} & Mark III {\it vs.} IRAS 1.2Jy & $ ?^\dagger$\\
 ITF \cite{dacosta98} & SFI {\it vs.} IRAS 1.2Jy & $ 0.6 \pm 0.1$\\
UMV \cite{zaroubi02b} & SEcat {\it vs.} PSCz & $ 0.51 \pm
 0.06$ \\
\hline
\end{tabular}
\end{center}
{\footnotesize \noindent $^\dagger$ Inconsistent flow fields (probably
due to problematic calibration of Mark~III)}
\end{table}

Figure~\ref{fig:zaroubi_fig3} shows the estimated value of $\beta$
from the PSC$z$ and the SEcat density-density comparison. The fields
where smoothed with $12 h{^{-1}}{\rm Mpc}$ Gaussian kernel; only
points with small errors, estimated from Monte Carlo mock SEcat
realizations, where used in the comparison. The left panel shows a
contour plot in the plane of $\beta$ and $\Delta\delta$ which
corresponds to zero-point density offset. The 1, 2 and 3-$\sigma$
level certainty contours are shown. These contours are drawn assuming
that the error in the density at each individual point is independent,
this assumption is obviously wrong. Therefore, in the $\beta$ results
quoted in the previous paragraph the uncertainty has been estimated by
adding, in quadrature, to the likelihood analysis 1-$\sigma$
uncertainties the error in recovering $\beta$ as estimated from the
Monte Carlo simulations. The right panel shows a scatter plot of
densities with their best fit linear model.  Figure
\ref{fig:zaroubi_fig4} shows results similar to the previous figure,
but for the reconstructed radial velocity field.  The comparison here
is somewhat different in the sense that it is carried out between the
reconstructed $12 h{^{-1}}{\rm Mpc}$ smoothed radial velocities at the
location of the SEcat data points. For more details on the comparison
see~\cite{zaroubi02b}.

Summary of the $\beta$ parameter values obtained from various
velocity-velocity and density-density comparisons is shown in
table~\ref{table:zaroubi_table2}.

Although, the results obtained by the UMV analysis strengthen the case
for lower $\beta$ values, by no account do they render the results of
the Mark~III-POTENT analysis invalid. However, they call for
re-examination of the later bearing in mind the following issues.  The
Mark~II catalog, as shown for example by Davis {\it et al.}
\cite{davis96} and more recently by Courteau {\it et al.}
\cite{courteau00}, suffers from systematic calibration error that
would cause a systematic error in the estimation of $\beta$. However,
this error is not expected to affect the density-density comparison as
to overestimate the value of $\beta$ by more than a factor of two. An
application of the UMV method to the Mark III
catalog~\cite{zaroubi02b} shows that the obtained values of $\beta$
are much lower than those obtained by applying the POTENT
analysis~\cite{sigad98} but still they are somewhat higher than those
obtained from the SEcat catalogs by $0.1-0.2$. Moreover, the
velocity-velcotiy analysis (like VELMOD) yield consistent values of
$\beta$ when applied to Mark~III and SFI datasets~\cite{branchini01,
willick98, willick97b}. Based on these arguments, one could speculate
that the most likely explanation to the previous inconsistency is a
result of a collusion between the systematic errors in Mark~III and
the noise sensitivity of the POTENT reconstruction.

\section{Summary \& Discussion}
\label{sec:zaroubi_summary}

To evaluate the outstanding issues in the peculiar velocity
measurements in general terms one should differentiate between two
types of problems. The first is the inconsistencies among results
obtained from various peculiar velocity data sets and occasionally
even within those obtained from the same data set. The second, is the
disagreement with measurements from other types of data sets. The
former is obviously more severe as it implies unresolved systematics
in the peculiar velocity datasets themselves.  Measuring by this yard
stick, the conflicting results obtained from the bulk flow and the
$\beta$ parameter measurements are more serious than the higher
amplitude found with the power spectrum measurements.


It is reassuring that within the local universe ($R \lsimZaroubi 60
h{^{-1}}{\rm Mpc}$) all the recent bulk flow measurements, especially from the
SFI, Shellflow and ENEAR catalogs do agree very well with each other
both in terms of amplitude and direction. Those three marginally agree
with the Mark~III data, which gives a somewhat higher value of $V_b$.
Given that the Shellflow sample have clearly shown that the Mark~III
data set is somewhat miscalibrated, this disagreement is hardly an
issue.

Unfortunately however, the picture on larger scales ($R\gsimZaroubi
100 h{^{-1}}{\rm Mpc}$) is very different and the disagreement among
the various measurements is yet to be resolved. While the bulk flow
obtained from the SNIa~\cite{riess97}, SCI/SCII~\cite{dale99} and
EFAR~\cite{colless01} samples clearly points toward convergence of the
CMB dipole; the SMAC\cite{hudson99}, LP \cite{lauer94} and LP10
\cite{willick99} surveys find a bulk flow amplitude of $\sim 600-700
{\rm km\,s{^{-1}}}$. Although the former three measurements are
consistent with the one obtained from the PSC$z$ redshift catalog
data, the NVSS radio sources and with theoretical prejudice, by no
means the later three have been refuted. It is worth pointing out
however, that the LP bulk flow, while comparable in amplitude,
disagrees with the SMAC and LP10 flow. Furthermore, almost all of the
deep peculiar velocity surveys have small number of objects and
therefore probably prone to the pitfalls of small number statistics.


The peculiar velocity measurements have systematically led to mass
power spectrum amplitudes higher than those obtained from other types
of data. In light of the overwhelming evidence pointing towards lower
amplitude power spectrum, a special effort should be made to rule out
any inherent bias in the {\it prior} assumptions made in the peculiar
velocity based power spectrum measurements. In fact, two recent
studies~\cite{hoffman00, silberman01} strongly suggest a generic
problem with the theoretical framework used to estimate the mass power
spectrum from the Mark~III, SFI, and ENEAR velocity surveys.  Possible
sources of this problem lie with non-linear dynamical effects and/or
oversimplified treatment of errors~\cite{silberman01}.


The irreconcilability of the $\beta$ estimation from density-density
and velocity-velocity comparisons has been one of the major
outstanding issues in the cosmic flows field of study. A new
technique, the UMV, have enabled carrying out, for the first time,
both comparisons in the same framework. The results obtained by
applying this technique to several galaxy peculiar velocity catalogs
yield low values of the $\beta$ parameter ($\sim 0.5-0.6$), a result
consistent with those obtained from the previous velocity-velocity
comparisons and from the analysis of redshift surveys.  These latest
results clearly strengthen the case for low $\beta$ values, in
agreement with those obtained by the previous velocity-velocity
studies. A qualitative inspection into the reason of the discrepancy
in the estimation of $\beta$ between the UMV and POTENT
density-density comparisons allows one to speculate that the most
likely explanation is a collusion between both the systematic errors
in the Mark~III data and the noise effects on the POTENT algorithm,
that somehow conspired to produce these high $\beta$-values.

In light of the developments presented in this review, it is argued
that most of the outstanding issues in the large scale peculiar
velocity field of study, {\it maybe} with the exception of the large
scale velocity field, have been either resolved or understood (at
least qualitatively) and we finally have a consistent cosmological
model emerging from the study of cosmic flows.  The experience gained
during the convergence towards the current status, despite the crooked
path it took, will be invaluable when dealing with the large future
datasets.

\bigskip

\noindent {\Large{\bf Acknowledgements}}

\bigskip

 Much of the work presented here have been done in collaboration with
E. Branchini, L.N. da Costa and Y. Hoffman, their contribution is
acknowledged.  I would like to thank the organizers of this meeting
for suggesting to me to review this topic.

\end{document}